\documentclass[aps, preprint, showpacs]{revtex4}
\begin{document}
\title{Topological Photon}
\author { S. C. Tiwari\\
Institute of Natural Philosophy\\
c/o 1 Kusum Kutir Mahamanapuri,Varanasi 221005, India}
\begin{abstract}
 We associate intrinsic energy equal to $h\nu/2$ with the spin angular momentum of photon and
propose a topological model based on orbifold in space and tifold in time as topological
obstructions. The model is substantiated using vector wavefield disclinations. The physical 
photon is suggested to be a particle like topological photon and a propagating wave such that 
the energy $h\nu$ of photon is equally divided between spin energy and translational energy
corresponding to linear momentum of $h\nu/c$. The enigma of wave-particle duality finds
natural resolution and the proposed model gives new insights into the phenomena of interference
and emission of radiation.
\end{abstract}
\maketitle

Introduction:-The emergence of quantum information science in recent years \cite{1} has led to a paradigm shift on the foundations of quantum mechanics: now counter-intuitive features of quantum theory are treated as powerful resources for the quantum information revolution. Technology to produce single photons on demand \cite{2} make quantum optics as one of the important avenues to realize this goal. In particular, angular momentum (AM)  both spin (SAM) and orbital (OAM) of photons has received a great deal of attention \cite{3}. In the process fundamental questions have also arisen on the nature of AM of photon. What is the meaning of 'intrinsic spin' of photon? Is OAM of photon intrinsic? Could one convert SAM to OAM? While these questions have been discussed in the literature it is baffling that the problem of energy associated with spin has escaped attention \cite{4}. Note that spin energy also remains obscure in classical electromagnetism.

	It is well known that Einstein introduced the directed linear momentum of ${h\nu}/c$ to the Planck's quantum oscillator of energy $h\nu$. Measurement of the spin of photon by Beth in 1935 should have led to the modification in the (quantum) harmonic oscillator picture of radiation; however even today photon in quantum optics \cite{5} retains essentially the same representation: 'The discrete excitations or quanta of the electromagnetic field, corresponding to the occupation numbers $\{n\}$, are usually known as photons'. Here $\{n\}$ denotes the Fock state of the radiation field. Is it possible to go beyond the undecidability \cite{6} of the physical reality of photon? Can one refute the extreme anti-photon view \cite{7} advocated by Lamb ? In this paper we present a definite model of photon recognizing that unlike energy and momentum, the spin of photon has a kind of topological invariance. The main hypothesis of topological photon is enunciated and elucidated. The idea of disclination in vector fields is shown to support novel topological construction. Finally it is pointed out that our model not only nicely resolves the age-old problem of wave-particle duality but also offers possible new photon physics.

Topological photon:- Past efforts to develop photon models have adopted the reductionist approach: from macroscopic electromagnetic fields to microscopic field quantum. In a comprehensive work 
\cite{8} we have argued that this approach has serious limitations, and instead of this it would be more fruitful and logical to develop a photon fluid theory for electromagnetic fields from first principles based on photons. It is imperative that for such a programme to succeed there must  exist  a viable photon model. In our approach geometry and topology of space and time are endowed physical reality \cite{8} and photon is visualized as extended space-time structure. A vortex model of photon was investigated recently articulating topological interpretation for spin in analogy with the quantized superfluid vortex \cite{9}. This model does not take into account the internal time periodicity and the postulated spin energy for the photon. A novel idea introduced here accomplishes this objective; we state it in the form of a hypothesis.

Hypothesis: Photon is a propagating topological defect in space and time characterized by the indices 1 and $1/2$ respectively.

	To substantiate the hypothesis we explain the nature of topological obstructions. Note that 4-dimensional space-time continuum having the metric with signature $(+, +, +, -)$ implies an asymmetry between space and time coordinates. In nature there also exist irreversible physical phenomena, therefore, we are led to a (3+1)-dimensional space and time. Without taking any recourse to field description the vortex of \cite{9} is now proposed to be a topological obstruction in space and identified as orbifold (orbital manifold) of Wilson-Sommerfeld (W-S) quantization. Photon spin is a topological invariant with index 1 given by a de Rham period integral (to be explained shortly) of 1-form in 3-dimension
\begin{equation}
 \oint {a}=1
\end{equation}

Internal time and associated energy is a subtle issue. In our model internal time periodicity is in harmony with the time period of space-translation of the object, therefore, both unidirectional and periodic time enter the picture. Internal time would imply closed time loops on 1-dimensional directed line of time identified as topological obstruction in time with the integral
\begin{equation}
 \oint {a_0} =1/2
\end{equation}
Plausible argument for index 1/2  can be given considering a 1-dimensional directed line such that the origin O is removed, and the transformation from left to right is a jump of  twice the interval. Eq.(2) embodies discreteness of time or 'a hole' in time continuum. Note the existence of kinks as point defects in 1-dimensional Ising model in condensed matter physics \cite{10} which bear resemblance with the proposed defects in time; let us call them tifolds. Thus a topological photon (TP) is orbifold+tifold.

	To understand the significance of de Rham's theorem we refer to an ingenious approach due to Post \cite{11} termed as quantum cohomology. He points out that Einstein in 1917 anticipated de Rham's theorem in his analysis of W-S condition
\begin{equation}
\oint{{p_i} {dq^i}} =n h
\end{equation}
Since momentum ${\bf p}$ is a gradient of Hamilton's scalar function the integral (3) would vanish for a Bohr's circular orbit. Einstein replaces the circular orbit by a torus and the momentum vector is made single valued with the integration loop comprising two sheets to obtain (3). The interior of torus is a topological obstruction in space (orbifold) in Einstein's analysis. In the language of differential forms the 1-form ${p_i}{dq^i}$ is closed but not exact, and the W-S integral is a de Rham period integral.

	Differential forms and topology are familiar in quantum optics literature since the discovery of geometric phases, therefore, a brief overview of the essentials would suffice and we refer to \cite{12} for  details. In a simple intuitive way a scalar function is a 0-form and its exterior derivative (d) gives a 1-form A (similar to gradient in 3-dimension) and dA is a 2-form (like curl of a vector). Stokes theorem in vector calculus relates line integral with a surface integral. Notice that line is a boundary of surface. Generalize it to forms: integration is carried out over chains; for a (p+1)-chain $C^{p+1}$ there is defined a boundary operator 
${C^p} = \partial{C^{p+1}}$. Homology and cohomology refer to the sets of chains and forms on a manifold respectively. A form $\omega$ is closed if $d\omega = 0$, and exact if $\omega =d\alpha$. Since ${d^2}=0$, $\omega=d\alpha$ implies $d\omega=0$ but $d\omega=0$ does not imply $\omega=d\alpha$. Similarly a chain C is called a cycle if $\partial C=0$, and boundary if $C=\partial B$. The integral $\oint\omega$ of a closed form over a cycle C is called a period integral of this form. In a given manifold the quotient space $H_p$ of the spaces of cycles ($Z_p$) and boundaries ($B_p$) of dimension p is defined the p-homology space. The cohomology space $H^p$ dual to $H_p$ is $Z^p/{B^p}$ where $Z^p(B^p)$ denotes the space of closed (exact) forms. Usually the groups $H_p$ and $H^p$ are said to be isomorphic however Post has drawn attention to the important but little known fact that isomorphism holds if topological torsion does not exist. Using Alexander duality \cite{13} he shows that torsion can exist only in dimension 4 or more; on the other hand Kiehn's definition \cite{14} based on a 3-form. $A\wedge  F = A\wedge  dA$ admits torsion in 3-dimension . We prefer Post's definition and ask if tifold could be viewed as torsion. Post links torsion with time as an asymmetry between homology and cohomology groups in 4-dimensional space-time. It would seem that this suggestion implies that torsion in time is contingent on the space periodicity; hence it cannot be tifold but would correspond to the time periodicity in the translation of TP in the external space. If a defect or obstruction in time is fundamental then tifold is a real topological object. To relate topological invariants with the physical property of photon using Planck's cpnstant h in Eq. (1) we get spin from orbifold and ‘internal’ energy from tifold i.e. Eq. (2)
\begin{equation}
E_i = h\nu/2
\end{equation}										

Here $\nu = \frac{1}{\tau}$, $\tau$ being a descrete time interval. If the time period of translational motion of TP is in harmony with $\tau$ then the total energy of photon is given by
\begin{equation}
E=E_i +E_l =h\nu
\end{equation}
										
In vacuum, photon energy is equally divided into internal energy and translational energy of the propagating defect; recall that a propagating defect is a wave \cite{15}. TP has a particle attribute and propagating discontinuity has wave attributes of wave vector and frequency; photon is wave plus particle. Note that topological invariance embodied in Eqs. (1) and (2) supersedes relativisitic invariance.

Field theory:-Singularities in fields obeying partial differential equations (usually of second order) are quite often indicative of topological defects. Heuristic analogy with the dislocation and disclination in crystals and singular vector wave fields has proved very useful to understand light phenomena. We seek a field theoretic description of the topological model adopting- 1) a conservative approach. No new field quantities are proposed and none of the electromagnetic potentials is superfluous; the question that intrigued Dirac in his failed attempt to give a new electron theory \cite{16}, and 2) a radical departure postulating electric $(\bf{E})$ and magnetic $(\bf{B})$ fields to be zero for single photon, and interpreting the Lorentz gauge condition as energy conservation law. Electromagnetic potentials (pure gauge) are proposed to represent the energy-momentum of the space-time fluid \cite{8}; it is easy to verify that electronic charge unit e can be factored out from the Maxwell field equations rendering the fields in geometrical units, and multiplication by Planck's constant gives the dimension of energy-momentum to the potentials. Thus the vector potential is redefined as $\hbar\bf{A}/e$. We have the following definitions
\begin{equation}
\bf{E}=-\nabla\Phi -\frac{\partial\bf{A}}{c\partial t} 
\end{equation}
\begin{equation}
\bf{B}=\nabla\times\bf{A}
\end{equation}									
and the Lorentz gauge condition is
\begin{equation}
\nabla .\bf{A}+\frac{\partial\Phi}{c\partial t}=0
\end{equation}

Setting	$\bf{E} = \bf{B} = 0$ Eqs. (6)-(8) can be combined to show that $(\bf{A},\Phi)$ satisfy the wave equation. In our earlier work \cite{9} fluid approach was used to construct vortex photon model; here new insight is gained based on pure disclination solution given by Nye \cite{17}. Disclination in a vector wave is a generalization of a dislocation in scalar wave \cite{18}. Let us consider monochromatic complex scalar wave field $\rho (r) exp[i\phi (r) - i\omega t]$ then dislocation is defined to be a phase singularity where $\rho = 0$ and $\phi (r)$ is indeterminate. Screw dislocation is a point defect with integer topological index (or an integral multiple of 
$2\pi$ phase) and helical wave front \cite{18,19}. We take a different perspective and ask whether phase singularity in 1-dimensional time for the phase $\omega t$ could be envisaged. A remarkable result is obtained in the following validating our tifold hypothesis in terms of the pure screw disclination in vector wave field.

	Instead of electric field vector and divergence equation for it in the work of Nye 
\cite{17} here we conider the vector $\bf {A}$ and Eq. (8). Much of the mathematical analysis is same as given by Nye: define z-axis as the direction of propagation and the transverse field 
$\bf{A}_t$ to lie in the xy-plane. Disclination is the locus of points where $A_x=A_y=0$. The direction of $\bf{A}_t$ on the moving line (disclination) is indeterminate, and a pure screw disclination using cylindrical coordinates $(r,\theta,z)$ is given by
\begin{equation}
A_x =k r e^{i\chi}
\end{equation}
\begin{equation}
A_y = i k r e^{i\chi}
\end{equation}
Note that for a typical scalar field such a solution could be imagined as a vortex around z-axis with uniform flow along it. Substitution of $A_x$ and $A_y$ in Eq. (8) gives
\begin{equation}
\frac {\partial A_z}{\partial z} +\frac{\partial\Phi}{c\partial t} =0
\end{equation}

Here we have defined $\chi =\theta + kz -\omega t$. Eq. (11) admits nonzero $A_z$ unlike $E_z = 0$ in Nye's analysis due to divergence equation for $\bf{E}$ used by him. Obviously $A_z$will give linear momentum and the relation $\omega = kc$ for a propagating wave. Nye considers the azimuth of transverse field in the xy plane
\begin{equation}
\beta=-\chi +2 n \pi
\end{equation}
and interprets the case with $z=0$ as a rigid rotation of the whole pattern with angular velocity 
$\frac{\omega}{2}$. For fixed t the pattern twists about z-axis by $\pi$ in one wavelength. Let us re-examine $z=0$ case for which
\begin{equation}
\beta - \theta = -2(\theta - \frac {\omega t}{2}) +2 n \pi
\end{equation}
Endowing topological defect in time itself, the phase condition (13) is equivalent to the         tifold with index $\frac {1}{2}$ for the term $\frac {\omega t}{2}$. New perspective on screw disclination for vector field $\bf{A}$ gives field theoretic demonstration of the topological hypothesis: transverse field $\bf{A}_t$ defines spin (dislocation or vortex), screw disclination of phase change $\pi$ is a tifold, and Eq. (11) gives the propagation law for the defects i.e. the topological photon.

Discussion and conclusion:-The first important result of our investigation is an elegant resolution of the age-old problem of wave-particle duality: TP has particle-attribute and propagating discontinuity is a wave. This picture is nearer in spirit to de Broglie’s original idea of wave and particle (also favored by Einstein) \cite{20} but discarding the description based on point particle and instantaneous time; finite discrete time and spatial extension of TP distinguish our model radically. Not only this the linear momentum is intrinsic to wave, therefore, in the phenomena like interference and diffraction the core particulate aspect i.e. TP does not play any role until its detection. There is a vast literature on the interference phenomena, however an important experiment reported in 1993 \cite{21} appears to support our interpretation. In this Young's double slit experiment using trapped ions as slits the scattered light is made to interfere. Authors argue that the observed pattern could be explained without invoking the position-momentum uncertainty relation. More important in the present context is the polarization-sensitive detection. Mercury ion used in the experiment has doubly degenerate excited state corresponding to magnetic quantum number $m_J$. Linearly polarized light scattered with $\Delta m_J=0$ is $\pi$-polarized and ion's state remains same while with $|\Delta m_J|=1$ the scattered light is 
$\sigma$-polarized and the ion's internal state undergoes a change. Interference exists for the former case while it disappears for the $\sigma$-polarized light. To quote from \cite{21} :"In this context, the existence of interference fringes indicates wavelike behavior, while the absence of fringes, consistent with a single photon trajectory, which begins at the source, intersects one of the atoms, and continues to the detector, indicates a particle like behavior". Note that in our model emission and absorption processes necessarily involve TP. Experimentally photon's momentum could show varied features independent of photon spin; recent experiments on recoil momentum with ultracold atoms do indicate interesting questions \cite{22}.

	The second result is on the zero-point energy: Eq. (1) implies topological origin of this energy. There are many experiments, e.g. Lamb shift, Casimir force etc. which demonstrate the effect of enigmatic quantum vacuum energy. There also exists Planck's 1912 ensemble based derivation of this energy, see \cite{11}. Topological zero-point energy proposed here is   associated with each photon's spin, and unlike mysterious quantum vacuum physical photons always possess this energy. Its implication on the black-body radiation is that the  Planck's formula represents the gas of photo-molecules i.e. the spin correlated photon pairs with AM zero \cite{4}. In the light of the importance of Planck’s radiation law from quantum physics to cosmology (microwave background radiation) \cite{23} the present interpretation assumes significance as a testable consequence of our model. Moreover the quantum vacuum of quantum field theories would be a kind of frozen solid phase of photons with zero momenta i.e. only topological photons. Besides these speculations, on a more practical side this approach offers the possibility of developing photon-fluid theory for the macroscopic electromagnetic fields and Maxwell equations.

	In conclusion, a novel topological photon model is proposed and its field theoretic validity is demonstrated based on vector wave disclinations.

I am grateful to Prof. E. J. Post who made me realize that I had been dwelling deep
into the world of topology without being aware of it, and educating me on cohomology. Useful correspondence on topological torsion with Professors R. M. Kiehn and E. Scholz is gratefully acknowledged. Library facility at Banaras Hindu University is acknowledged.

\end{document}